\documentclass{elsart}
\usepackage{amsmath}
\usepackage{tipa}
\usepackage{amssymb}
\usepackage{multirow,setspace,times,amssymb,amsmath,graphicx,color,rotating,subfigure,url}
\usepackage{lineno}
\usepackage[square,sort&compress,comma]{natbib}

\journal{Physica A} 

\begin{document}

\begin{frontmatter}

\title{Long-term correlations and multifractal analysis of trading volumes for Chinese stocks}
\author[BS,SS,BUTE]{Guo-Hua Mu},
\author[SZSE]{Wei Chen},
\author[BUTE,HUT]{J{\'a}nos Kert{\'e}sz},
\ead{kertesz@phy.bme.hu} %
\author[BS,SS,RCE,RCSE,RCFE]{Wei-Xing Zhou\corauthref{cor}}
\ead{wxzhou@ecust.edu.cn} %
\corauth[cor]{Corresponding author. Address: 130 Meilong Road, P.O.
Box 114, School of Business, East China University of Science and
Technology, Shanghai 200237, China, Phone: +86 21 64253634, Fax: +86
21 64253152.}

\address[BS]{School of Business, East China University of Science and Technology, Shanghai 200237, China}
\address[SS]{School of Science, East China University of Science and Technology, Shanghai 200237, China}
\address[BUTE]{Department of Theoretical Physics, Budapest University of Technology and Economics, Budapest, Hungary} %
\address[SZSE]{Shenzhen Stock Exchange, 5045 Shennan East Road, Shenzhen 518010, China}
\address[HUT]{Laboratory of Computational Engineering, Helsinki University of Technology, Espoo, Finland}%
\address[RCE]{Research Center for Econophysics, East China University of Science and Technology, Shanghai 200237, China}
\address[RCSE]{Engineering Research Center of Process Systems Engineering (Ministry of Education), East China University of Science and Technology, Shanghai 200237, China}
\address[RCFE]{Research Center on Fictitious Economics \& Data Science, Chinese Academy of Sciences, Beijing 100080, China}

\begin{abstract}
We investigate the temporal correlations and multifractal nature of
trading volume of 22 liquid stocks traded on the Shenzhen Stock
Exchange in 2003. We find that the trading volume exhibits
size-dependent non-universal long memory and multifractal nature. No
crossover in the power-law dependence of the detrended fluctuation
functions is observed. Our results show that the intraday pattern in
the trading volume has negligible impact on the long memory and
multifractality.
\end{abstract}


\end{frontmatter}

\section{Introduction}
\label{S1:Intro}

As implied by a well-known adage in the Wall Street that it takes
volume to move stock prices, trading volume contains much
information about the dynamics of price formation. For instance, the
investigation of price-volume relationship has a long history in
finance \cite{Karpoff-1987-JFQA}, attracting more and more interest
of physicists, and recently has been studied at the transaction
level
\cite{Chan-Fong-2000-JFE,Lillo-Farmer-Mantegna-2003-Nature,Lim-Coggins-2005-QF,Naes-Skjeltorp-2006-JFinM,Zhou-2007-XXX}.
In addition, the distributions of trading volumes at different time
scales have been found to have power-law right tails for different
stock markets
\cite{Gopikrishnan-Plerou-Gabaix-Stanley-2000-PRE,Eisler-Kertesz-2006-EPJB,Eisler-Kertesz-2007-PA,Queiros-2005-EPL,deSouza-Moyano-Queiros-2006-EPJB},
which can account at least partly the power-law tails of returns
\cite{Gabaix-Gopikrishnan-Plerou-Stanley-2006-QJE,Gabaix-Gopikrishnan-Plerou-Stanley-2007-JEEA,Gabaix-Gopikrishnan-Plerou-Stanley-2008-JEDC,Zhou-2007-XXX}.
The distributions of trading volumes for Chinese stocks have also
been reported to have power-law tails
\cite{Mu-Chen-Kertesz-Zhou-2009-EPJB,Qiu-Zhong-Chen-2009-PA,Zhou-2007-XXX}.

Another important feature of trading volumes is its long-range
temporal correlation. Lobato and Velasco used a two-step
semiparametric estimator in the frequency domain for the long-memory
parameter $d$ of daily trading volume for 30 stocks composing DJIA
from 1962 to 1994 and find that $d=0.30\pm0.08$
\cite{Lobato-Velasco-2000-JBES}, which amount to the Hurst index
$H=d+0.5=0.80\pm0.08$. Gopikrishnan et al performed detrended
fluctuation analyses of trading volume for 1000 largest US stocks
over the two-year period 1994-1995
\cite{Gopikrishnan-Plerou-Gabaix-Stanley-2000-PRE}. They found that,
the trading volumes at different time scales (from 15 min to 390
min) show stronger correlations with $H=0.83\pm0.02$. Bertram used
the autocorrelation and variance plots to investigate the memory
effect of high-frequency trading volumes for 200 most actively
traded stocks on the Australian Stock Exchange spanning the period
January 1993 - July 2002, and found that the average Hurst index is
$H=0.79\pm0.03$ \cite{Bertram-2004-PA}. Qiu et al conducted similar
analysis on 18 liquid Chinese stocks from 2004 to 2006 and reported
that $H=0.83$, which does not depend on the intraday pattern
\cite{Qiu-Zhong-Chen-2009-PA}.

By investigating the TAQ data sets of 2674 stocks in the period
2000-2002, Eisler and Kert\'{e}sz found that the strength of
correlations depends on the liquidity of stocks so that the Hurst
index increases logarithmically with the average trading volume or
the company size
\cite{Eisler-Kertesz-2006-EPJB,Eisler-Kertesz-2006-PRE,Eisler-Kertesz-2006,Eisler-Kertesz-2007-PA,Eisler-Bartos-Kertesz-2008-AP}.
There is also evidence showing that trading volumes possess
multifractal nature in different markets, such as the high-frequency
trading volumes of 30 DJIA constituent stocks
\cite{Moyano-Souza-Queiros-2006-PA}, of New York Stock Exchange
stocks \cite{Eisler-Kertesz-2007-EPL} and of the Korean stock index
KOSPI \cite{Lee-Lee-2007-PA}. These properties have not been studied
for the Chinese market, which will be investigated in this work.

The paper is organized as follows. We give a brief description of
the data in Section \ref{S1:Data}. The intraday pattern, temporal
correlations and multifractal nature of trading volume are studied
in Section \ref{S1:Results}, where we will show that the intraday
pattern has negligible impact on the temporal correlations and
multifractal nature. Section \ref{S1:conclusion} concludes.

\section{Data sets}
\label{S1:Data}

We analyze an ultra-high-frequency database containing 22 Chinese
stocks traded on the Shenzhen Stock Exchange in 2003. It records the
sizes of all individual transactions. The 22 stocks investigated in
this work cover a variety of industry sectors such as financials,
real estate, conglomerates, metals \& nonmetals, electronics,
utilities, IT, transportation, petrochemicals, paper \& printing and
manufacturing. Our sample stocks were part of the 40 constituent
stocks included in the Shenzhen Stock Exchange Component Index in
2003
\cite{Zhou-2007-XXX,Gu-Chen-Zhou-2007-EPJB,Mu-Chen-Kertesz-Zhou-2009-EPJB}.
The market started with an opening call auction period from 9:15
A.M. to 9:25 A.M. followed by a 5-min cooling period, and then the
market entered the double continuous auction period
\cite{Zhou-2007-XXX,Gu-Chen-Zhou-2007-EPJB,Mu-Chen-Kertesz-Zhou-2009-EPJB}.
We focus on the trades occurred in the double continuous auction
period.

The tickers of the 22 stocks investigated are the following: 000001
(Shenzhen Development Bank Co. Ltd), 000002 (China Vanke Co. Ltd),
000009 (China Baoan Group Co. Ltd), 000012 (CSG holding Co. Ltd),
000016 (Konka Group Co. Ltd), 000021 (Shenzhen Kaifa Technology Co.
Ltd), 000024 (China Merchants Property Development Co. Ltd), 000027
(Shenzhen Energy Investment Co. Ltd), 000063 (ZTE Corporation),
000066 (Great Wall Technology Co. Ltd), 000088 (Shenzhen Yan Tian
Port Holdings Co. Ltd), 000089 (Shenzhen Airport Co. Ltd), 000429
(Jiangxi Ganyue Expressway Co. Ltd), 000488 (Shandong Chenming Paper
Group Co. Ltd), 000539 (Guangdong Electric Power Development Co.
Ltd), 000541 (Foshan Electrical and Lighting Co. Ltd), 000550
(Jiangling Motors Co. Ltd), 000581 (Weifu High-Technology Co. Ltd),
000625 (Chongqing Changan Automobile Co. Ltd), 000709 (Tangshan Iron
and Steel Co. Ltd), 000720 (Shandong Luneng Taishan Cable Co. Ltd),
and 000778 (Xinxing Ductile Iron Pipes Co. Ltd).

\begin{figure}[htb]
\centering
\includegraphics[width=7cm]{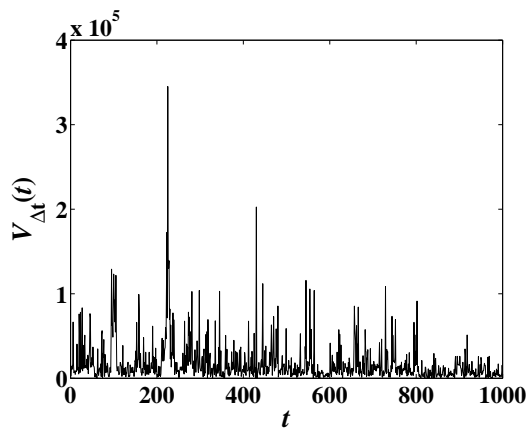}
\caption{\label{Fig:ExampleVol} A segment of the time series of
1-min trading volume $V_{\Delta{t}}$ for stock 000001.}
\end{figure}

Let $v_i$ be the size of the $i$-th trade for a given stock and $N
\equiv N_{\Delta t}$ the number of trades in a fixed time interval
$\Delta t$. Then the trading volume at time scale $\Delta{t}$ is
\begin{equation}
 V_{\Delta t} = \sum_{i=1}^{N}{v_{i}}.
 \label{Eq:V}
\end{equation}
A segment of the time series of 1-min trading volume $V_{\Delta{t}}$
for stock 000001 is illustrated in Fig.~\ref{Fig:ExampleVol}.

\section{Results}
\label{S1:Results}

\subsection{Intraday pattern}
\label{S2:IntraPattern}

\begin{figure}[htb]
\centering
\includegraphics[width=6.5cm]{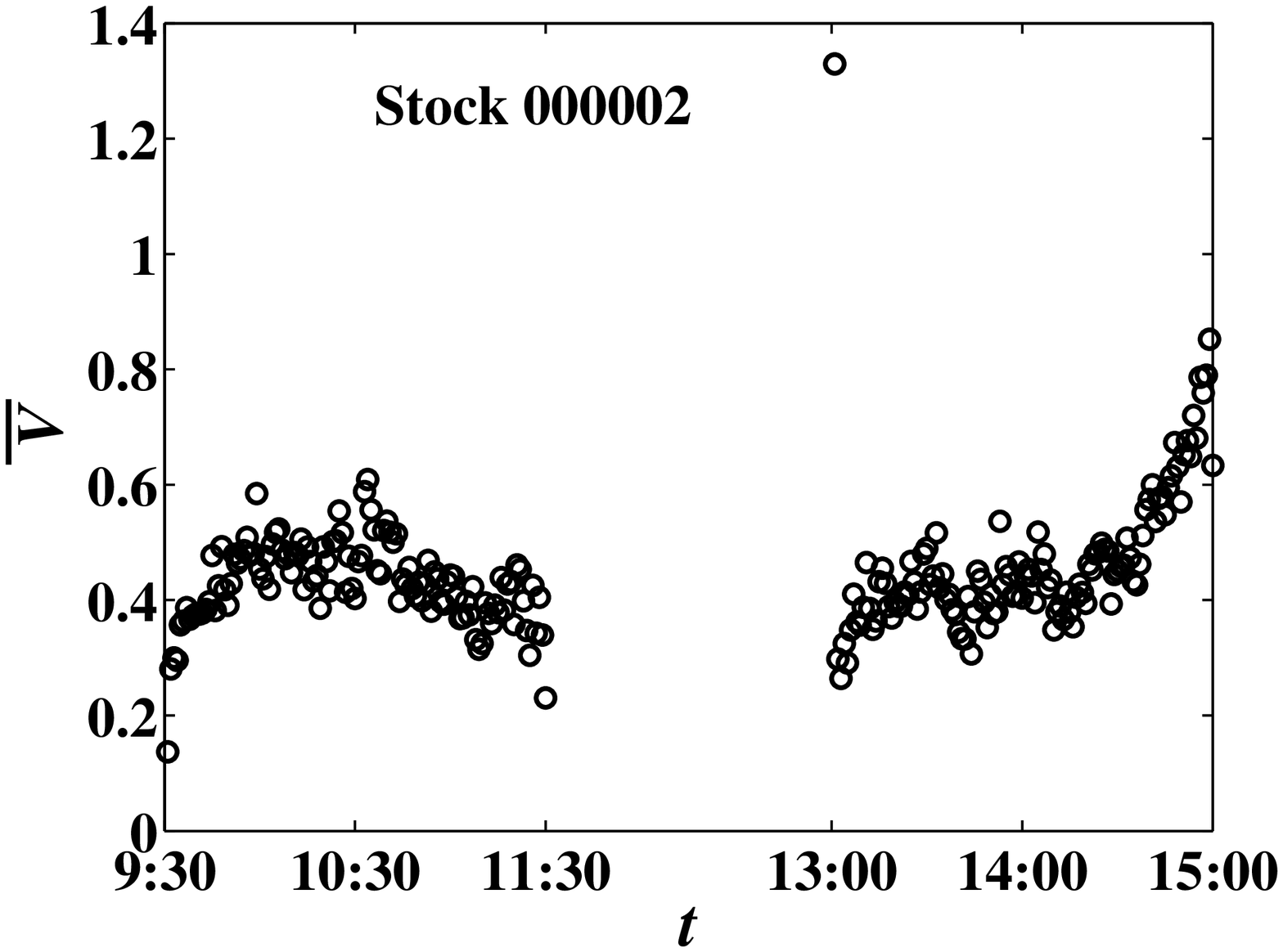}
\includegraphics[width=6.5cm]{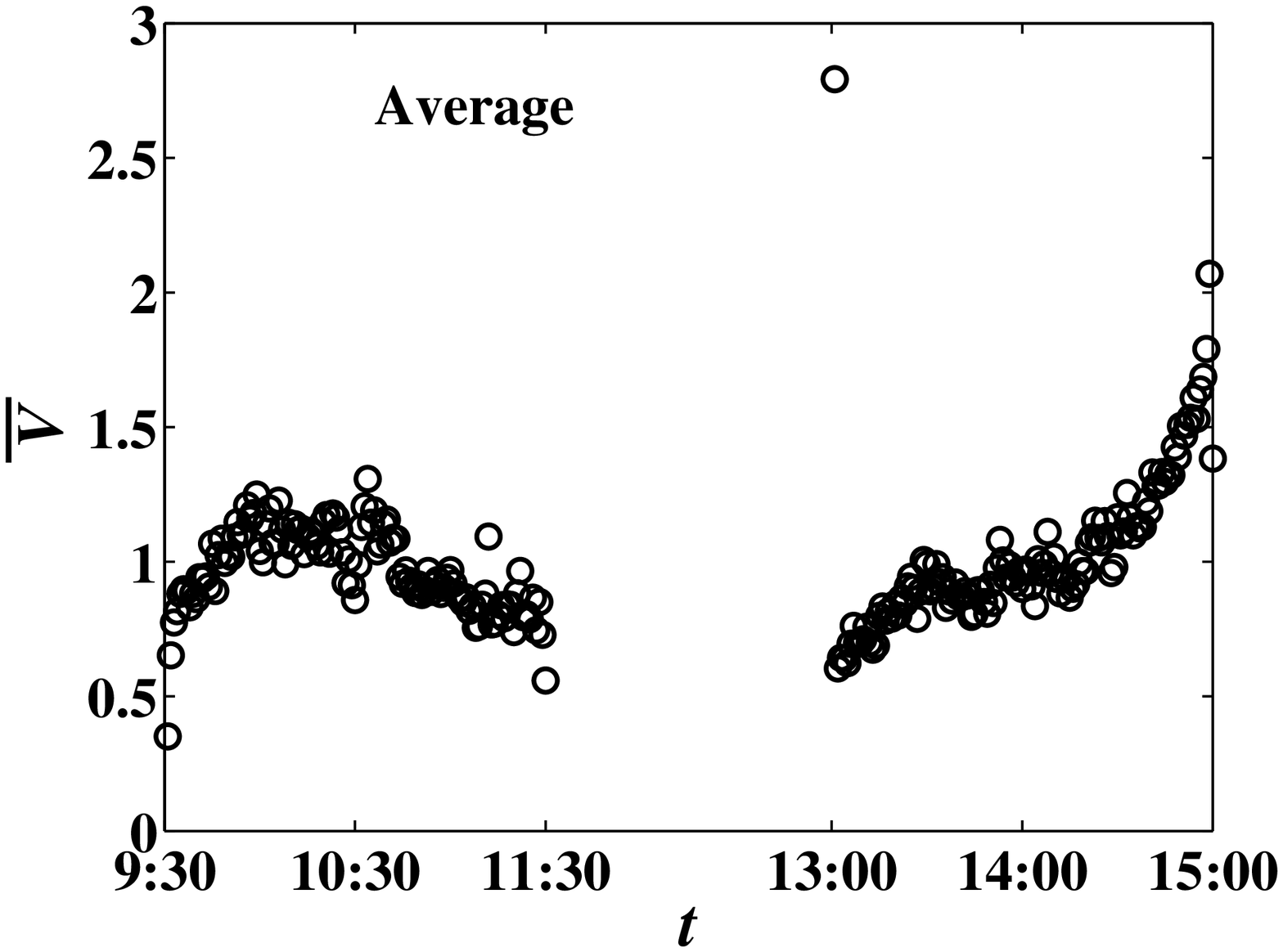}
\caption{\label{Fig:IntraPattern} Intraday pattern of trading
volumes is given. Stock 000002 (code number) is shown as an example
of individual stock in the left plot, while average result of 22
stocks is listed in the right.}
\end{figure}

Many researches report that there exist intraday patterns in the
trading volume  but with different shapes
\cite{Wood-McInish-Ord-1985-JF,Admati-Pfleiderer-1988-RFS,Stephan-Whaley-1990-JF,Gopikrishnan-Plerou-Gabaix-Stanley-2000-PRE,Lee-Fok-Liu-2001-JBFA,Qiu-Zhong-Chen-2009-PA}.
Figure \ref{Fig:IntraPattern} gives the intraday pattern of trading
volume for stock 000002 and the average for all 22 stocks. The
average trading volume increases and then decreases in the morning,
with two mild peaks at about 10:00 and 11:00. At the first minute in
the afternoon, there is a significant jump, which is simply due to
the fact that orders submitted in the noon closing period $(11:30 -
13:00)$ are executed at 13:00. Afterwards, we see a monotonic
increase in the average trading volume. Our result is quite similar
to that in Refs.
\cite{Lee-Fok-Liu-2001-JBFA,Qiu-Zhong-Chen-2009-PA}, and the
intraday pattern does not has a U-shape.

\subsection{Size-dependent correlation in trading volume}
\label{S2:H:lnV}

We investigate the temporal correlations of trading volumes based on
the detrended fluctuation analysis (DFA)
\cite{Peng-Buldyrev-Havlin-Simons-Stanley-Goldberger-1994-PRE,Kantelhardt-Bunde-Rego-Havlin-Bunde-2001-PA}
which is a special case of the multifractal DFA method
\cite{Kantelhardt-Zschiegner-Bunde-Havlin-Bunde-Stanley-2002-PA}. If
the time series are long-range power-law correlated, the detrended
fluctuation function $F_{q}(s)$ versus $s$ could be describe as
follow,
\begin{equation}
 F_{q}(s)={\langle [F^2(s)]^{q/2} \rangle}^{1/q} \sim s^{h(q)},
 \label{Eq:MFDFA}
\end{equation}
where $s$ is the length of each segments (time window size) and
$F^2(s)$ is the variance of the detrended time series in a given
segment after removing a linear trend, while $h(q)$ is the
generalized Hurst exponent. When $q=2$, we have
\begin{equation}
 F_{2}(s)\sim s^{H},
 \label{Eq:DFA}
\end{equation}
which gives the well-known Hurst exponent $H$.

\begin{figure}[htb]
\centering
\includegraphics[width=6.5cm]{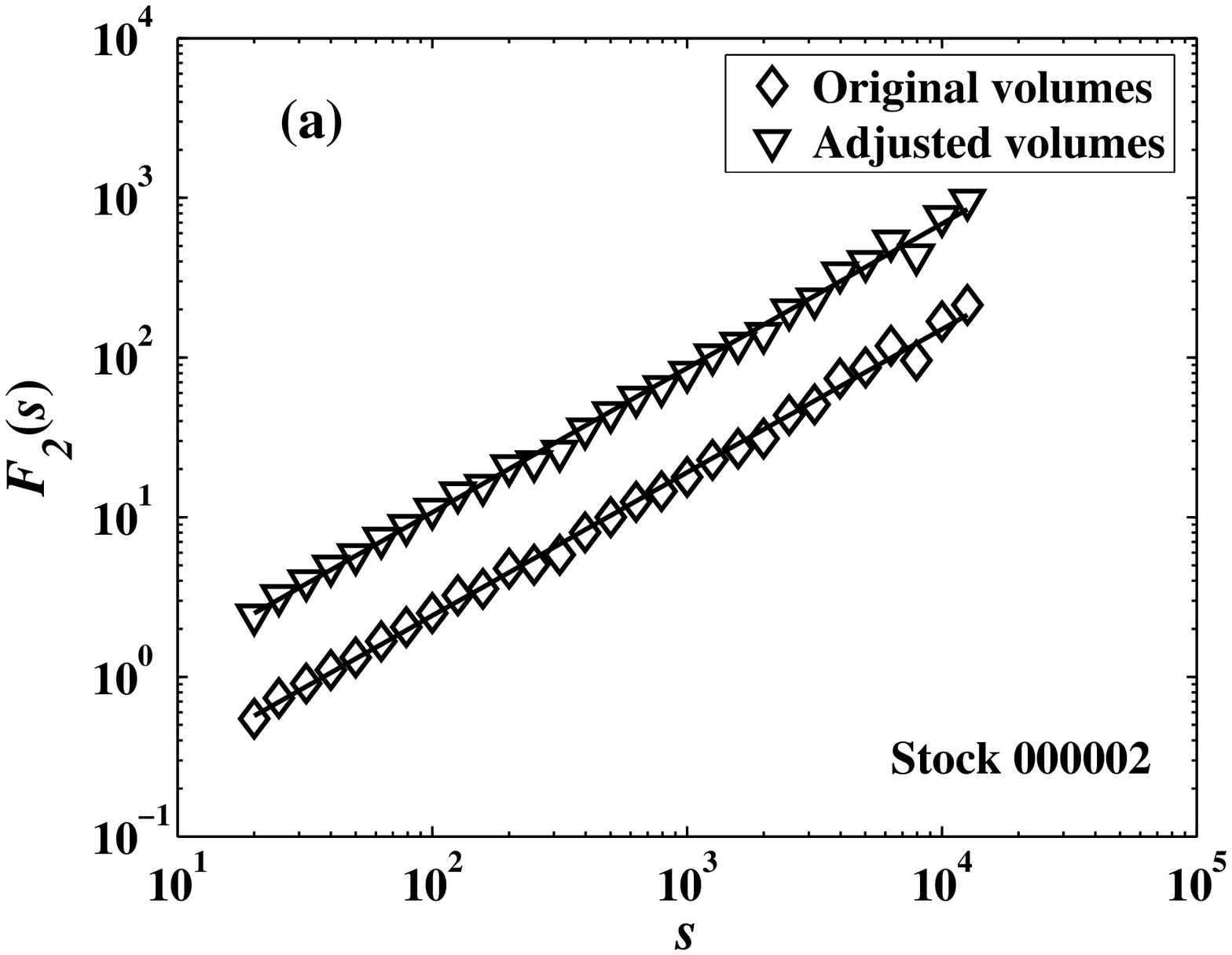}
\includegraphics[width=6.5cm]{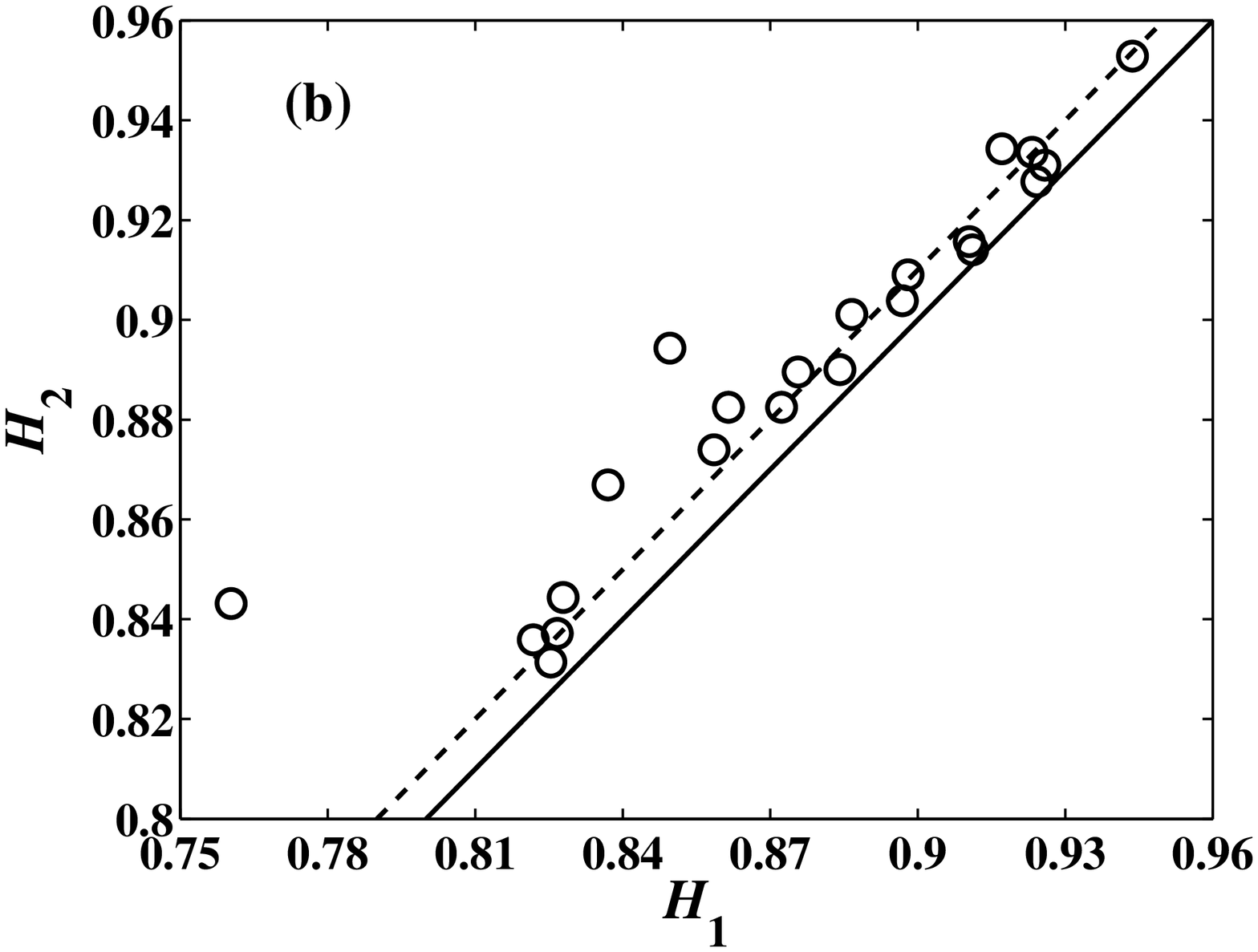}
\caption{\label{Fig:DFA} Detrended fluctuation analysis of 1-min
trading volume. (a) Dependence of $F_{2}(s)$ on $s$ for original and
deseasonalized trading volumes for stock 000002. (b) Relationship
between the Hurst indexes $H_1$ and $H_2$ of original and
deseasonalized trading volumes. The solid line is $H_2=H_1$ and the
dashed line is $H_2=H_1+0.01$.}
\end{figure}

We perform DFA on the 1-min original trading volume data and the
deseasonalized (or adjusted) data after removing the intraday
pattern for each stock. Fig. \ref{Fig:DFA}(a) illustrates the
power-law dependence of $F_{2}(s)$ on $s$ for stock 000002. In
addition, the two curves are almost parallel, indicating that the
intraday pattern has negligible impact on the memory effect of
trading volume. The results are similar for other stocks. The slopes
of the best fitted linear lines in Fig.~\ref{Fig:DFA}(a) give the
estimates of the Hurst indexes for the original data ($H_1$) and the
adjusted data ($H_2$), which are presented in Table \ref{Tb:VolDFA}.
The average Hurst indexes are $\bar{H_1}=0.88\pm0.05$ and
$\bar{H_2}=0.89\pm0.04$. We also plot $H_2$ against $H_1$ in
Fig.~\ref{Fig:DFA}(b) to show that $H_2\approx H_1$. A careful
scrutiny shows that $H_2$ is slightly greater than or equal to
$H_1$. We note that there is no crossover in the DFA plot of trading
volume for Chinese stocks, which should be compared to the fact that
there is no consensus for the presence of crossover
\cite{Gopikrishnan-Plerou-Gabaix-Stanley-2000-PRE,Eisler-Kertesz-2007-PA}.

\begin{figure}[htb]
\centering
\includegraphics[width=6.5cm]{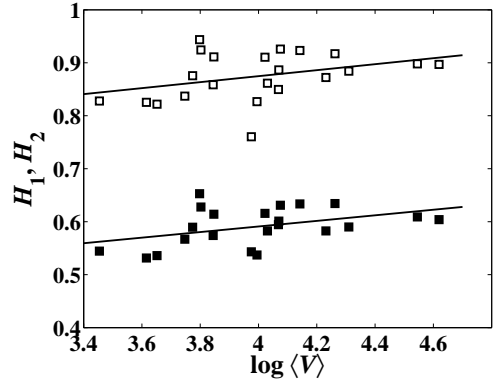}
\caption{\label{Fig:H:lnV} Logarithmic dependence of Hurst index on
average trading volume for the original data ($\square$) and
adjusted data ($\blacksquare$). The data points for the adjusted
trading values have been shifted vertically downwards for better
visibility.}
\end{figure}

It is important to stress that the Hurst index varies from one stock
to another, which depends on the company size or average trading
volume of a stock. In Fig. \ref{Fig:H:lnV}, we present the
dependence of the Hurst indexes of trading volumes on different
logarithmic values of $\langle V \rangle$, the sample average of
$V_{\Delta{t}}$ for individual stocks. We find that there is a
linear relationship for both original and deseasonalized data:
\begin{equation}
 H_i=H_i^*+\gamma_{H_i}\log\langle V\rangle,~~~~~~i=1,2
 \label{Eq:H:lnV}
\end{equation}
where the base of log is 10, $\gamma_{H_1}=0.06\pm0.03$ for the
original data, and $\gamma_{H_2}=0.05\pm0.03$ for the adjusted data.
The relation (\ref{Eq:H:lnV}) was first observed by Eisler and
Kert\'{e}sz for the traded value (also called dollar volume or
capital flow, defined by the trading volume times stock price), and
they found that $\gamma_H=0.06\pm0.01$ for NYSE stocks and
$\gamma_H=0.05\pm0.01$ for NASDAQ stocks
\cite{Eisler-Kertesz-2006-PRE,Eisler-Kertesz-2006-EPJB,Eisler-Kertesz-2007-PA}.
Jiang et al verified the relationship for the traded values of about
1500 Chinese stocks and obtained that $\gamma_H=0.013\pm0.001$
\cite{Jiang-Guo-Zhou-2007-EPJB}. Since large average trading volume
corresponds roughly to large company size (or capitalization), our
results show that larger company has stronger correlation in the
trading volume. We note that the relation (\ref{Eq:H:lnV}) is
observed for the first time for trading volume in this work.

\subsection{Mean-variance analysis}
\label{S2:H:dt}

\begin{figure}[htb]
\centering
\includegraphics[width=6.5cm]{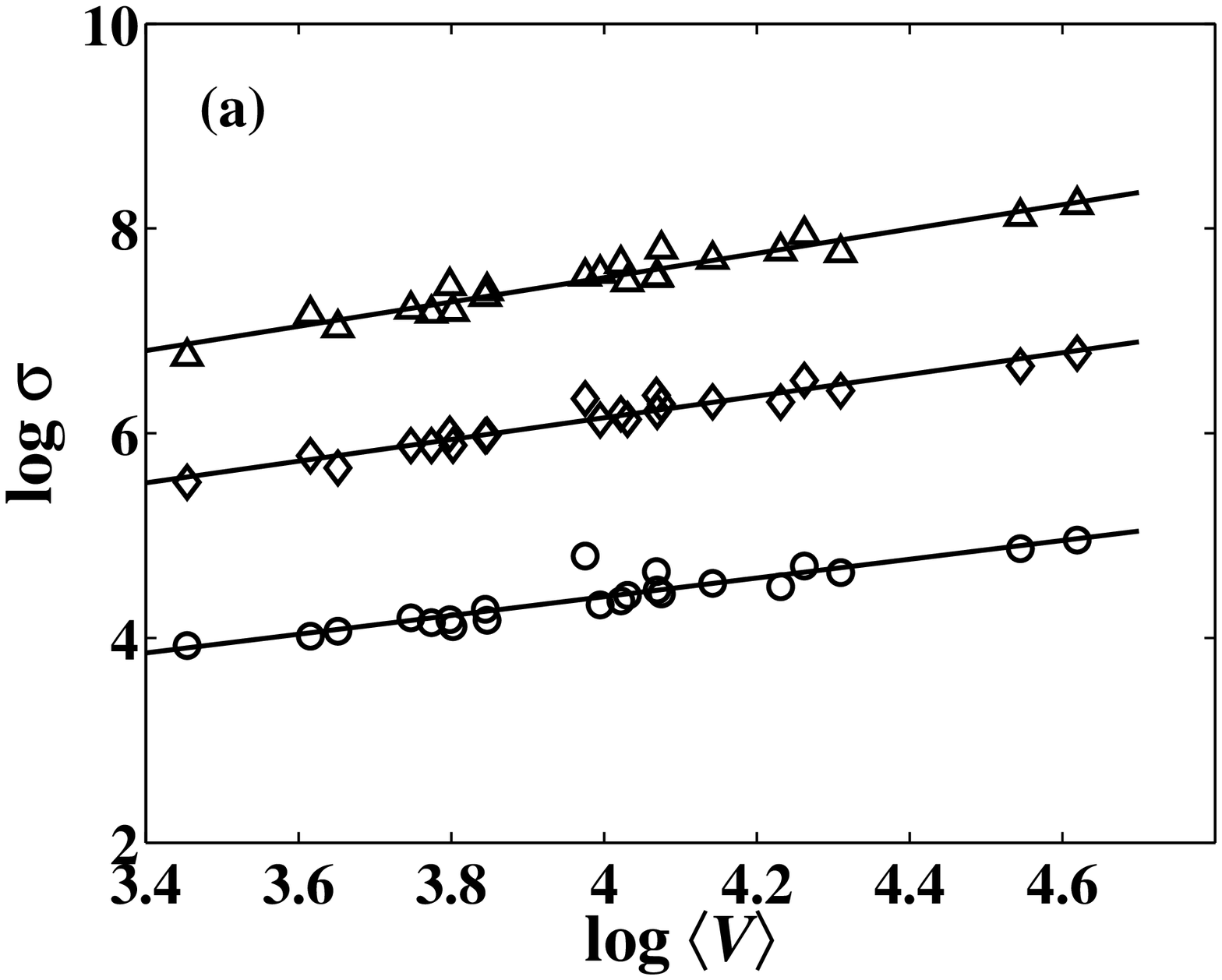}
\includegraphics[width=6.5cm]{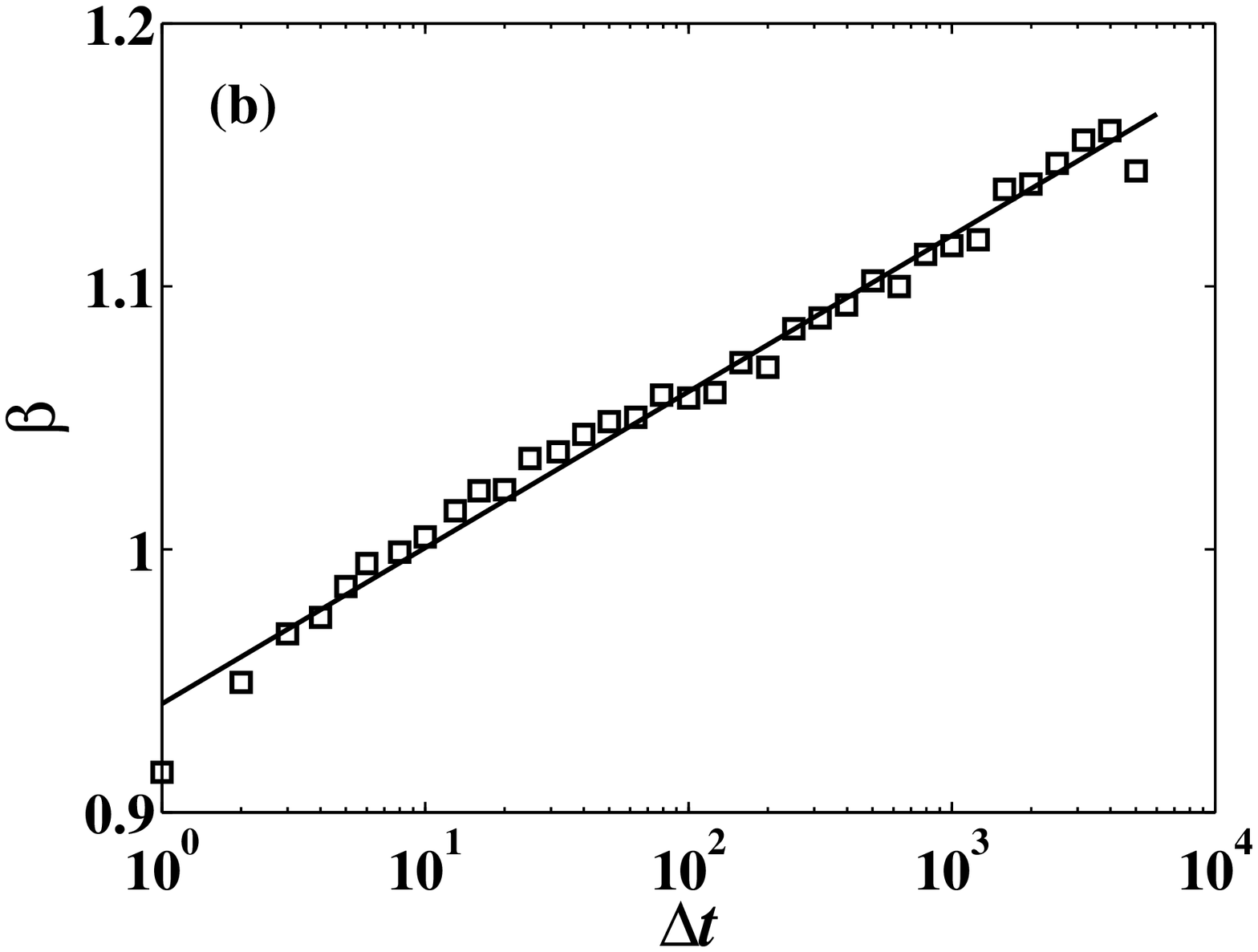}
\caption{\label{Fig:mean:var} Mean-variance analysis of trading
volume. (a) Power-law dependence of $\sigma_{\Delta{t}}$ on
$\langle{V_{\Delta{t}}}\rangle$ for $\Delta t=$1 min ($\circ$), 120
min ($\lozenge$) and 20 days ($\vartriangle$) for the original data.
(b) Logarithmic dependence of the scaling exponent $\beta$ on the
time scale $\Delta t$ for the original data. }
\end{figure}

We now conduct the mean-variance analysis on the time series of
trading volume. For each stock, the mean $\langle V_{\Delta{t}}
\rangle$ and the variance $\sigma_{\Delta{t}}^2$ are calculated for
different time scales $\Delta{t}$. Since $V_{\Delta{t}}$ is
additive, the mean-variance analysis gives \cite{Taylor-1961-Nature}
\begin{equation}
 \sigma_{\Delta{t}} \sim \langle V_{\Delta{t}} \rangle ^{\beta}~,
 \label{Eq:Fluctuation_V}
\end{equation}
where $\langle \cdot \rangle$ denotes time averaging.
Fig.~\ref{Fig:mean:var}(a) illustrates the power-law dependence of
$\sigma_{\Delta{t}}$  with respect to $\langle V_{\Delta{t}}
\rangle$ in double logarithmic coordinates for three time windows
$\Delta t=$1 min, 0.5 trading day (120 min) and 20 trading days and
for the original data. For the deseasonalized data,
$\langle{V}\rangle\equiv1$ so that the mean-variance does not apply.
The slopes of the best linear fits give the estimates of $\beta$ at
different time scales $\Delta{t}$. Fig.~\ref{Fig:mean:var}(b) plots
$\beta$ as a function of $\Delta{t}$ for the original trading volume
data, which has a logarithmic trend,
\begin{equation}
\beta=\beta^*+\gamma_{\beta}\log\Delta{t},
 \label{Eq:Bdependence}
\end{equation}
where $\gamma_{\beta}=0.059\pm0.001$. We find that the following
relation holds
\begin{equation}
\gamma_{\beta} \approx \gamma_{H_i},~~~~~~i=1,2,
 \label{Eq:slope}
\end{equation}
which has been well verified for the traded values for different
stock markets including developed
\cite{Eisler-Kertesz-2006-EPJB,Eisler-Kertesz-2006-PRE} and emerging
stock markets \cite{Jiang-Guo-Zhou-2007-EPJB}.

\subsection{Multifractal analysis} \label{S2:MF}

In this section, we employ the MF-DFA method
\cite{Kantelhardt-Zschiegner-Bunde-Havlin-Bunde-Stanley-2002-PA} to
investigate the multifractal nature of trading volumes. In this
procedure, the $q$-th order fluctuation function $F_{q}(s)$ versus
$s$ is analyzed for different $q$. We vary the value of $s$ in the
range from $s_{\min}=20$ to $s_{\max}=M/4$ ($M$ is the length of a
series), since $F_{q}$ becomes statistically unreliable for very
large scales $s$, and systematic deviations will be involved for
very small scales $s$. The relationship between $h(q)$ and the mass
scaling exponents $\tau(q)$ in the conventional multifractal
formalism based on the partition functions
\cite{Halsey-Jensen-Kadanoff-Procaccia-Shraiman-1986-PRA,Kantelhardt-Zschiegner-Bunde-Havlin-Bunde-Stanley-2002-PA}
is formalized as follows,
\begin{equation}
 \tau(q)=qh(q)-D_{f}~,
 \label{Eq:tau}
\end{equation}
where $D_{f}$ is the fractal dimension of the geometric support of
the multifractal measure (in our case $D_{f}$ = 1). According to the
Legendre
transform~\cite{Halsey-Jensen-Kadanoff-Procaccia-Shraiman-1986-PRA},
we have
\begin{equation} \label{Eq.f2}
 \alpha = h(q)+qh'(q)~~~~ \mbox{and}~~~~
 f(\alpha) = q(\alpha-h(q))+1,
\end{equation}
providing the estimation of strength of singularity $\alpha$ and its
spectrum $f(\alpha)$.

\begin{figure}[htb]
\centering
\includegraphics[width=6.5cm]{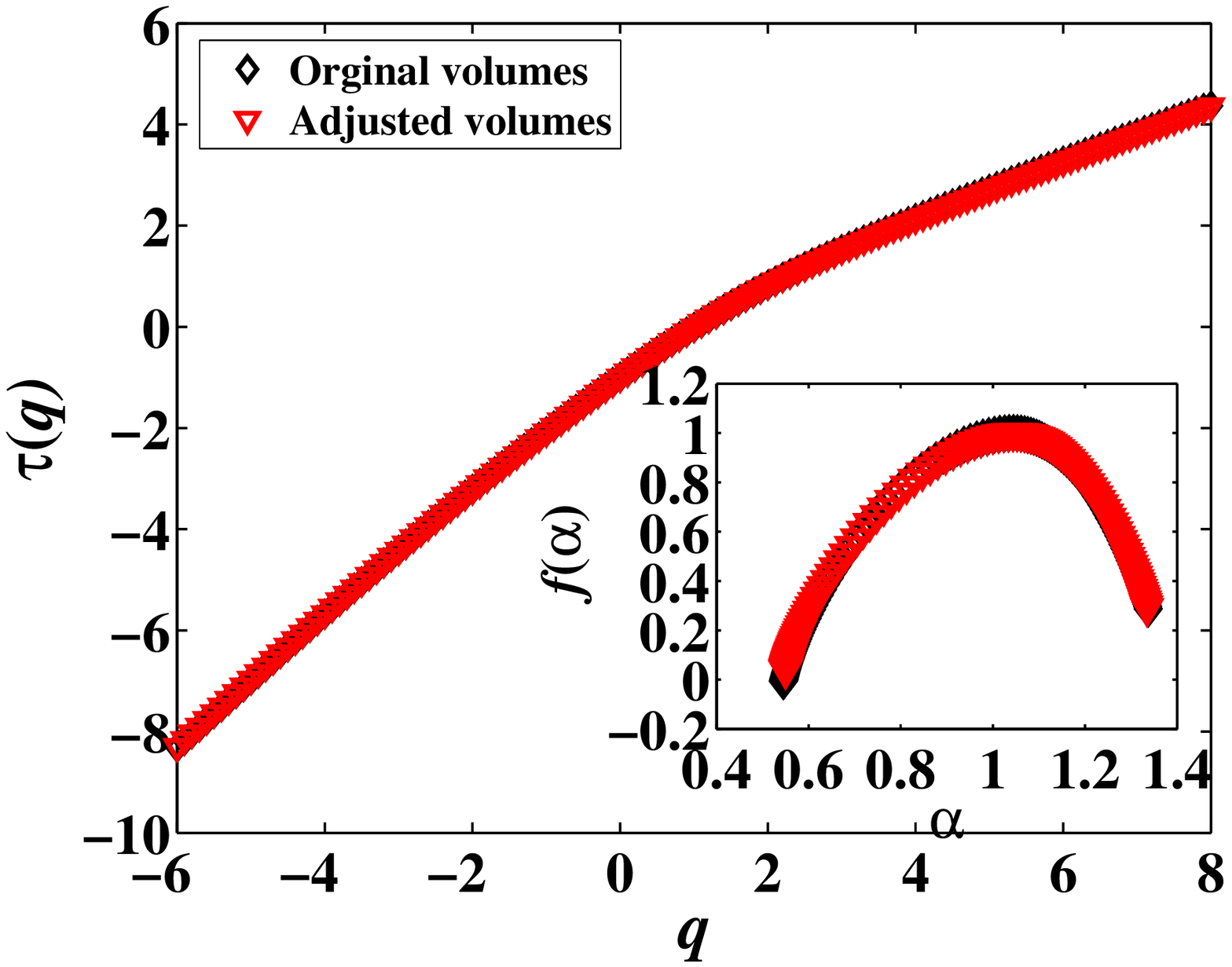}
\includegraphics[width=6.5cm]{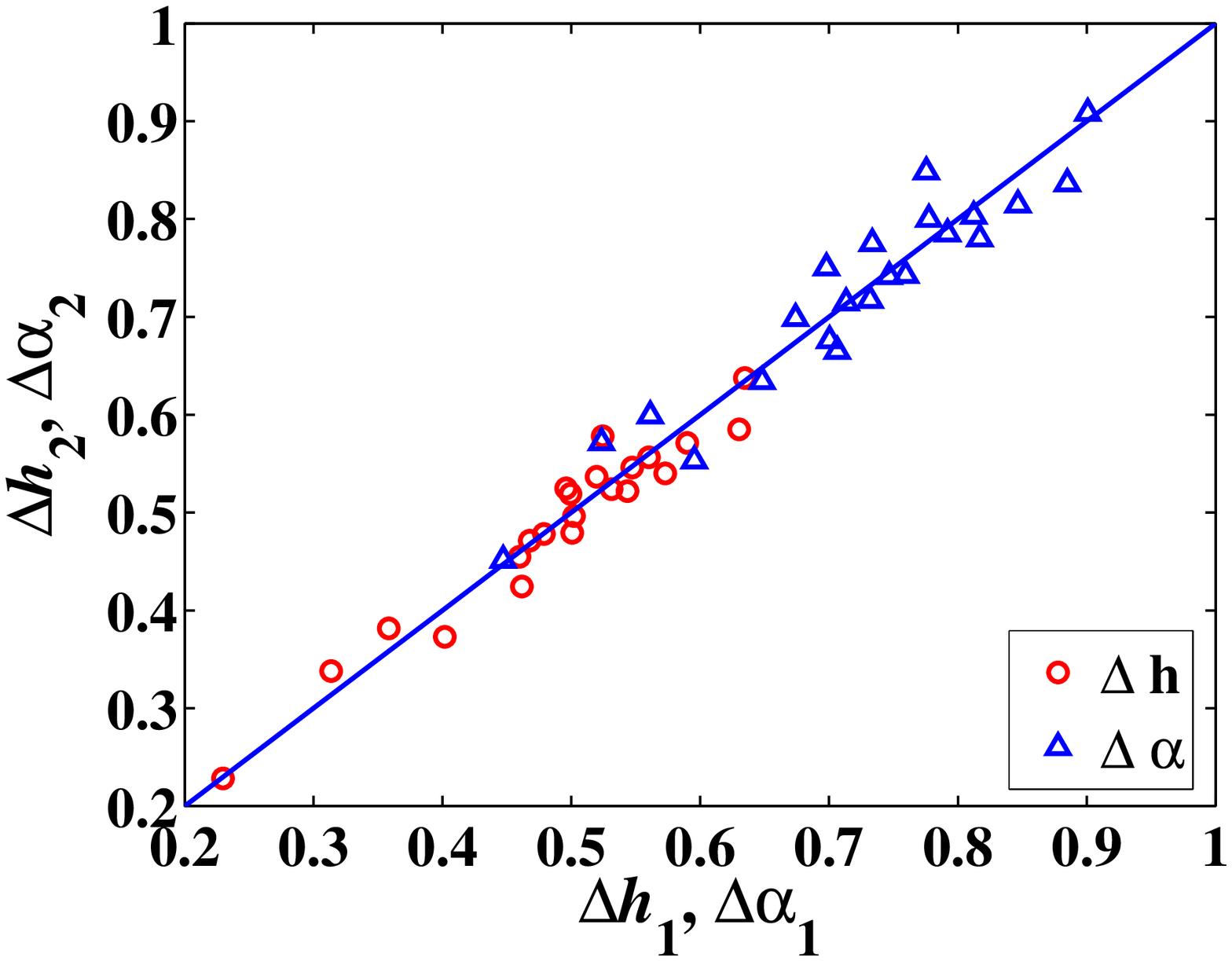}
\caption{\label{Fig:MFDFA} Multifractal detrended fluctuation
analysis of trading volumes for stock 000002. (a) Non-linear
dependence of $\tau(q)$ with respect to $q$ and the multifractal
spectrum $f(\alpha)$ in the inset. (b) Negligible impact of intraday
pattern on the multifractal nature of trading volumes.}
\end{figure}

Fig.~\ref{Fig:MFDFA} illustrates an example the multifractal
analysis for stock 000002. It is evidence from
Fig.~\ref{Fig:MFDFA}(a) that $\tau(q)$ is a non-linear function of
$q$, which is a hallmark for the presence of multifractality. There
is no remarked discrepancy observed between the original and
adjusted data of trading volumes. To further clarify the negligible
influence of the intraday pattern on the multifractal nature of
trading volumes, we calculate two characteristic values $\Delta{h} =
h_{\max}-h_{\min}$ and $\Delta \alpha=\alpha_{\max}-\alpha_{\min}$
respectively for each time series. The results of $\Delta h$ and
$\Delta \alpha$ are listed in Table~\ref{Tb:VolDFA}. The results are
presented in Fig.~\ref{Fig:MFDFA}(b). We see that
$\Delta{h_1}\approx\Delta{h_2}$ and
$\Delta{\alpha_1}\approx\Delta{\alpha_2}$, indicating that the
intraday pattern in the trading volume has negligible impact on the
multifractal nature.

\section{Conclusion}
\label{S1:conclusion}

We have studied the temporal correlations and multifractal nature of
trading volume for 22 most actively traded Chines stocks on the
Shenzhen Stock Exchange. Detrended fluctuation analysis shows that
the trading volumes at different time scales possess non-universal
long memory, whose Hurst index $H$ depends logarithmically on the
average trading volume as $H=H^*+\gamma_H\log\langle{V}\rangle$. The
mean-variance unveils that the scaling exponent $\beta$ depends
logarithmically on the time scale as
$\beta=\beta^*+\gamma_{\beta}\log\Delta{t}$. Empirical evidence
shows that $\gamma_H=\gamma_{\beta}$, consistent with the
theoretical derivation. The investigation of the size-dependent
non-universal correlation in trading volume has not been conducted
before. Multifractal detrended fluctuation analysis confirms that
the trading volume exhibits multifractal nature. Comparing the
results obtained from the original trading volume data and the
adjusted data after removing the intraday pattern, we conclude that
the intraday pattern has negligible impact on the temporal
correlations and multifractal nature of trading volumes.

In general, the results obtained for the Shenzhen Stock Exchange in
this paper are qualitatively the same as other emerging and
developed markets. However, there are some differences. There are
studies showing that there is a crossover phenomenon in the
power-law relation between the detrended fluctuation function and
the scale for some markets, which is not observed for the Shenzhen
Stock Exchange. Also, the multifractal spectra differ from one
market to another with different singularity width, which is usually
determined by the distribution and the correlation structure of the
time series showing the idiosyncracy of different markets.

\bigskip
{\textbf{Acknowledgments:}}

This work was partly supported by the National Natural Science
Foundation of China (70501011 and 70502007), the Shanghai
Educational Development Foundation (2008SG29), the China Scholarship
Council (2008674017), and the Program for New Century Excellent
Talents in University (NCET-07-0288).

\bibliography{E:/Papers/Auxiliary/Bibliography}

\newpage
\section{Appendix}

\begin{table}[htp]
 \caption{\label{Tb:VolDFA} Appendix table. The subscript ``1'' means original data and the subscript ``2'' stands for adjusted data.}
 \medskip
 \centering
 \begin{tabular}{ccccccccccccccc}
 \hline \hline
    Stock code && $H_1$ & $H_2$  && $\Delta{h}_1$ & $\Delta{h}_2$ && $\Delta{\alpha}_1$ & $\Delta{\alpha}_2$\\%
    \hline
    000001 && $0.90\pm0.015$ & $0.91\pm0.014$ && 0.40 & 0.37 && 0.60 & 0.55\\%
    000002 && $0.90\pm0.009$ & $0.90\pm0.009$ && 0.55 & 0.55 && 0.79 & 0.78\\%
    000009 && $0.88\pm0.009$ & $0.89\pm0.008$ && 0.56 & 0.56 && 0.81 & 0.80\\%
    000012 && $0.83\pm0.005$ & $0.84\pm0.005$ && 0.46 & 0.45 && 0.65 & 0.63\\%
    000016 && $0.92\pm0.006$ & $0.93\pm0.006$ && 0.36 & 0.38 && 0.56 & 0.60\\%
    000021 && $0.91\pm0.015$ & $0.92\pm0.015$ && 0.52 & 0.58 && 0.78 & 0.85\\%
    000024 && $0.88\pm0.005$ & $0.89\pm0.005$ && 0.50 & 0.52 && 0.70 & 0.75\\%
    000027 && $0.92\pm0.016$ & $0.93\pm0.016$ && 0.46 & 0.42 && 0.71 & 0.67\\%
    000063 && $0.89\pm0.007$ & $0.90\pm0.007$ && 0.57 & 0.54 && 0.82 & 0.78\\%
    000066 && $0.91\pm0.009$ & $0.91\pm0.009$ && 0.50 & 0.50 && 0.73 & 0.72\\%
    000088 && $0.82\pm0.007$ & $0.84\pm0.007$ && 0.54 & 0.52 && 0.76 & 0.74\\%
    000089 && $0.85\pm0.011$ & $0.89\pm0.008$ && 0.63 & 0.64 && 0.90 & 0.91\\%
    000429 && $0.94\pm0.014$ & $0.95\pm0.014$ && 0.23 & 0.23 && 0.45 & 0.45\\%
    000488 && $0.86\pm0.007$ & $0.88\pm0.007$ && 0.47 & 0.47 && 0.67 & 0.70\\%
    000539 && $0.76\pm0.007$ & $0.84\pm0.008$ && 0.63 & 0.59 && 0.88 & 0.84\\%
    000541 && $0.83\pm0.009$ & $0.84\pm0.009$ && 0.48 & 0.48 && 0.71 & 0.71\\%
    000550 && $0.93\pm0.017$ & $0.93\pm0.017$ && 0.50 & 0.52 && 0.73 & 0.78\\%
    000581 && $0.84\pm0.006$ & $0.87\pm0.006$ && 0.52 & 0.54 && 0.78 & 0.80\\%
    000625 && $0.87\pm0.012$ & $0.88\pm0.011$ && 0.59 & 0.57 && 0.85 & 0.81\\%
    000709 && $0.92\pm0.010$ & $0.93\pm0.010$ && 0.31 & 0.34 && 0.52 & 0.57\\%
    000720 && $0.83\pm0.007$ & $0.83\pm0.007$ && 0.50 & 0.48 && 0.70 & 0.68\\%
    000778 && $0.86\pm0.008$ & $0.87\pm0.007$ && 0.53 & 0.52 && 0.75 & 0.74\\%
    \hline\hline
 \end{tabular}
\end{table}

\end{document}